\documentclass[conference]{IEEEtran}
\IEEEoverridecommandlockouts
\usepackage{cite}
\usepackage{amsmath,amssymb,amsfonts}
\usepackage{algorithm}
\usepackage{algpseudocode}
\usepackage{multirow}
\usepackage{graphicx}
\usepackage{textcomp}
\usepackage{xcolor}
\usepackage{threeparttable}
\usepackage{tabularx}
\usepackage{booktabs}
\usepackage{array}

\def\BibTeX{{\rm B\kern-.05em{\sc i\kern-.025em b}\kern-.08em
    T\kern-.1667em\lower.7ex\hbox{E}\kern-.125emX}}
\begin{document}

\title{Efficient Bilinear Attention-based Fusion for Medical Visual Question Answering}
\author{\IEEEauthorblockN{Zhilin Zhang\textsuperscript{1}, Jie Wang\textsuperscript{3}, Zhanghao Qin\textsuperscript{2}, Ruiqi Zhu\textsuperscript{3} and Xiaoliang Gong\textsuperscript{3*}\thanks{\IEEEauthorrefmark{1}Corresponding author.}}
\IEEEauthorblockA{\textsuperscript{1}Tandon School of Engineering, New York University, New York, USA}
\IEEEauthorblockA{\textsuperscript{2}School of Electrical and Electronic Engineering, Nanyang Technological University, Singapore}
\IEEEauthorblockA{\textsuperscript{3}College of Electronic and Information Engineering, Tongji University, Shanghai, China}
\IEEEauthorblockA{zz10068@nyu.edu, ZHANGHAO001@e.ntu.edu.sg, \{2054310, zhurq, gxllshsh\}@tongj.edu.cn}
}
\maketitle

\begin{abstract}
Medical Visual Question Answering (MedVQA) has attracted growing interest at the intersection of medical image understanding and natural language processing for clinical applications. By interpreting medical images and providing precise answers to relevant clinical inquiries, MedVQA has the potential to support diagnostic decision-making and reduce workload across various fields like radiology. While recent approaches rely heavily on unified large pre-trained Visual-Language Models, research on more efficient fusion mechanisms remains relatively limited in this domain. In this paper, we introduce a fusion model, OMniBAN, that integrates \textit{O}rthogonality loss, \textit{M}ulti-head attention, and a \textit{B}ilinear \textit{A}ttention \textit{N}etwork to achieve high computational efficiency as well as solid performance. We conduct comprehensive experiments and demonstrate how bilinear attention fusion can approximate the performance of larger fusion models like cross-modal Transformer. Our results show that OMniBAN requires fewer parameters (approximately 2/3 of Transformer-based Co-Attention) and substantially lower FLOPs (approximately 1/4), while achieving comparable overall performance and even slight improvements on closed-ended questions on two key MedVQA benchmarks. This balance between efficiency and accuracy suggests that OMniBAN could be a viable option for real-world medical image question answering, where computational resources are often constrained.
\end{abstract}


\section{Introduction}
Medical Visual Question Answering (MedVQA) is an emerging field within multi-modal artificial intelligence that adapts the principles of general Visual Question Answering (VQA) to meet the specific demands of the medical domain. The primary goal of MedVQA is to support healthcare professionals by automatically generating accurate answers to clinical questions based on medical images, thereby assisting in clinical decision-making and relieving workload. This task involves the fusion of computer vision and natural language processing techniques to analyze visual data alongside natural language questions to enable contextually relevant and clinically accurate responses.

Compared to general domain Visual Question Answering, MedVQA presents unique and significant challenges. At the image level, medical images such as those in radiology or pathology often exhibit subtle differences within small regions, where even minor pixel variations can represent completely different diagnostic findings (e.g., tiny lesions). At the text level, clinical inquiries are characterized by specialized terminology and complex language structures. This necessitates language models equipped with domain-specific knowledge to accurately interpret the semantics of medical questions. Given these challenges, the multi-modal fusion module is important, as it must ensure the effective integration of information from these distinct modalities without losing key information.

Despite the recent success of multimodal fusion techniques in enhancing MedVQA performance, there is a significant gap in research focusing on computationally efficient fusion methods. Transformer-based models, especially cross-modal Transformers, have demonstrated strong fusion capabilities and have been widely adopted in this domain. However, these large unified models come with substantial computational demands, making them less suitable for real-time clinical applications or small MedVQA model training, where computational efficiency is essential. Also, a common training strategy for MedVQA involves directly utilizing the embeddings from frozen visual and textual encoders, therefore, the primary computational cost of training concentrates on the fusion network. This motivates the need for exploring alternative fusion techniques that maintain high performance while reducing computational complexity.

In this paper, we propose an efficient fusion framework called OMniBAN, which combines \textit{O}rthogonality loss, \textit{M}ulti-head attention, and a \textit{B}ilinear \textit{A}ttention \textit{N}etwork. Our approach is designed to deliver comparable performance on MedVQA tasks at a lower training and computational cost. Through extensive experiments, we show that OMniBAN achieves similar results compared with large Transformer-based fusion models on key MedVQA benchmarks, yet it requires fewer computational resources. This efficiency–accuracy trade-off positions OMniBAN as a promising solution for real-world medical image question answering, particularly in radiological Visual Question Answering.

\section{Related Work}

\begin{figure*}
\centering
\includegraphics[width=\textwidth]{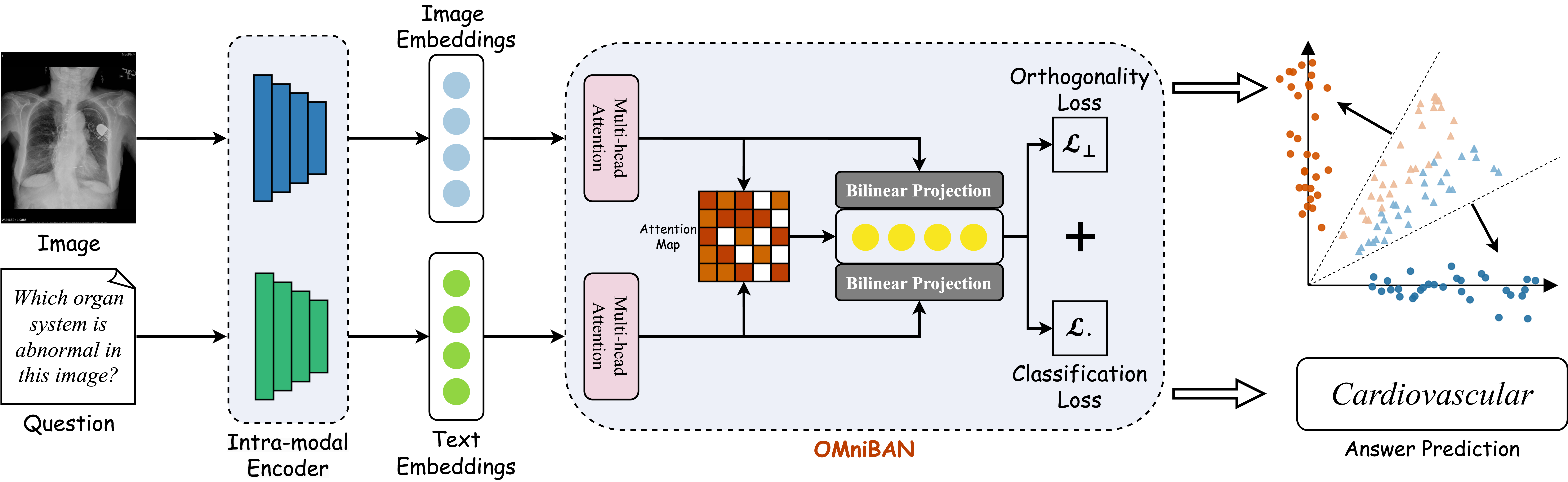}
\caption{Overview of our proposed Orthogonal Multi-head Bilinear Attention Network (OMniBAN). The frozen visual and textual backbones encode the input image and question independently, and then the core OMniBAN fusion module fuses these features using multi-head attention for within-modality refinement and bilinear attention to capture cross-modal interactions. Orthogonality loss is adopted to encourage diverse attention patterns among glimpses during training.}
\label{fig1}
\end{figure*}

\subsection{Medical Visual Question Answering}

Medical Visual Question Answering (MedVQA) is an emerging research area within multimodal artificial intelligence that applies the general principles of Visual Question Answering (VQA) to the medical domain. This field combines medical image understanding and natural language processing techniques to analyze and understand medical images in conjunction with natural language questions, with the goal of generating accurate answers useful for clinical decision-making and diagnostic assistance.

Initial research efforts in MedVQA adopt models that have proven effective in general VQA domain, and adapt them for medical applications. In terms of image feature extraction, researchers often rely on pre-trained models like VGGNet \cite{vggnet} and ResNet \cite{resnet}, which are fine-tuned for the specific task of MedVQA. On the text side, GRU \cite{chung2014empirical} and LSTM \cite{hochreiter1997long} are commonly used to extract textual features, while some approaches incorporate additional semantic information derived from medical corpora to enhance the embeddings used for question representation.

To address the challenges specific to MedVQA, such as the scarcity of labeled medical data, various techniques have been proposed. Nguyen et al. \cite{MEVF} introduced the Model-Agnostic Meta-Learning (MAML) framework combined with a Convolutional Denoising Auto-Encoder (CDAE) to improve feature learning. Similarly, Liu et al. \cite{CPRD} utilized contrastive learning to train a pre-trained model (CPRD) that was then applied to MedVQA tasks. These approaches often use transfer learning to leverage external datasets and pre-trained models to enhance the quality of the extracted image and text features.

\subsection{Multi-modal Fusion}
In the Visual Question Answering task, multi-modal fusion plays a crucial role by integrating visual and textual features to enable accurate classification. The performance of VQA models largely depends on the effective intra-modal feature extraction and the subsequent inter-modal fusion.

Recent approaches to multi-modal fusion in VQA have introduced various methods to enhance the integration of visual and textual features. A foundational method is SAN\cite{SAN}, which iteratively refines attention on relevant image regions based on the given question. Bilinear pooling has been a major focus in improving fusion by capturing complex interactions between modalities. Yu et al. \cite{MFB} proposed MFB to address the high computational cost of bilinear pooling by factorizing the interaction into two low-rank matrices, preserving performance while reducing complexity. MUTAN\cite{MUTAN} builds on this by applying Tucker decomposition to further compress the bilinear tensor and get a more compact and efficient multi-modal representation. To make full use of bilinear attention maps, Kim et al.\cite{BAN} proposed BAN, which can capture dependencies between visual and textual features efficiently.

The advent of Transformer\cite{transformer} has brought a significant shift in multi-modal fusion strategies. Methods like LXMERT\cite{lxmert}, hi-VQA\cite{radstruct}, MCAN\cite{MCAN} and METER\cite{METER} utilize cross-modality Transformers with separate encoders for vision and language, followed by a cross-modality encoder to fuse the extracted features. In the domain of Medical Visual Question Answering, Liu et al.\cite{q2a} employs a Transformer-based architecture that directly fuse image and text features to generate joint representation. While Transformers are effective for fusion, they require more computational resources and are more complex compared to bilinear pooling-based methods. This makes balancing performance and computational cost particularly important in Medical Visual Question Answering, where medical data is often limited.

\section{Method}\label{Method}

\subsection{Problem Formulation}\label{Problem Formulation}

Medical Visual Question Answering is regarded as a classification task, and the objective is to identify the most probable answer $a$ from a predefined set of possible answers $A = \{a_1, a_2, a_3, \dots, a_n\}$. This prediction can be expressed as:
\begin{equation}
\Hat{a} = \arg\max_{a \in A} P(a \mid v_i, q_i)
\end{equation}
where $ P(a \mid v_i, q_i) $ denotes the probability of a given answer $a$ being correct given the image $v_i$ and the question $q_i$ , and $\Hat{a}$ is the predicted answer that maximizes this probability.

\subsection{Multi-modal Feature Extraction}
\textbf{Image Encoder.} MedVQA requires highly specialized image encoders capable of capturing the intricate details specific to medical images, which often differ significantly from general image data. Medical imaging tasks demand a high level of precision, as even subtle visual cues can hold crucial clinical significance. 

 In this work, we employ the pre-trained BiomedCLIP Image Encoder\cite{biomedclip} rather than other methods as our image encoder. The advantage of BiomedCLIP are two-fold. First, BiomedCLIP builds on the CLIP model \cite{CLIP}, which was originally designed to learn image and text representations within a shared feature space through natural language supervision. This design has been proved strong zero-shot performance across various domains. Second, BiomedCLIP further fine-tunes CLIP on the PMC-15M dataset, which consists of diverse medical images and associated text, thereby improving its ability to handle medical visual data. 
 
 Given an input radiological image $ I_i \in \mathbb{R}^{H\times W\times C} $, BiomedCLIP first produces a hidden representation $v_{hid}$:
\begin{equation}
    v_{hid} = BiomedCLIP(I_i)
\end{equation}

This hidden representation is then passed through a projection layer to yield the final 512-dimensional image feature vector $v_i$ representing the initial visual features:
\begin{equation}
    v_i = Proj(v_{hid})
\end{equation}

\textbf{Question Encoder.} In this work, We adopt the pre-trained BioBERT model \cite{BioBERT} as our question encoder due to its suitability for processing complex biomedical language. BioBERT, based on the BERT architecture, was fine-tuned on a large and diverse biomedical corpus, and demonstrated strong ability in domain-specific biomedical language representation. Therefore, it can generate more accurate representations of medical questions than general-purpose language models.

Compared to traditional VQA models that often use recurrent neural networks, such as LSTM \cite{hochreiter1997long} and GRU \cite{chung2014empirical}, for text encoding, BioBERT and other BERT-based models provide notable advantages. The Transformer architecture in BERT is particularly effective at capturing long-range dependencies and contextual relationships within text, which can produce richer and more contextually accurate representations of questions. This capability is essential for handling the sophisticated language requirements of MedVQA, where accurate understanding of medical terminology and context is crucial. In our approach, BioBERT encodes each question as a 768-dimensional vector, denoted as $q_i$.

While BiomedCLIP also offers a text encoder, we chose not to use it for question encoding in this work. Although the BiomedCLIP Text Encoder aligns image and text features within a shared feature space, it lacks the word-level granularity necessary for capturing detailed linguistic information. This level of detail is critical for the fusion methods introduced in Section \ref{OMniBAN}, which benefit from precise, word-level representations.

\subsection{Orthogonal Multi-head Bilinear Attention Network}\label{OMniBAN}

Our proposed Orthogonal Multi-head Bilinear Attention Network (OMniBAN) integrates a single-layer multi-head self-attention mechanism with bilinear attention networks to efficiently fuse visual and textual features for Medical Visual Question Answering. This design can help capture complex intra-modal and cross-modal interactions effectively while maintaining computational efficiency. By leveraging orthogonal multi-head attention, OMniBAN enhances feature diversity and maximizes information extraction across modalities.

\subsubsection {Intra-modal Feature Attention}\label{intra-selfatt}

In OMniBAN, we employ a single layer of multi-head self-attention to act as intra-modal attention to refine image and question features independently before cross-modal fusion. Given image features $ v_i \in \mathbb{R}^{N_v \times d_v} $ and question features $ q_i \in \mathbb{R}^{N_q \times d_q} $, where $N_v=1$ for image features (since CLIP outputs global image-level features) and $N_q$ denotes the sequence length for question features, we apply multi-head self-attention separately to each modality. 
\begin{itemize}
\item \textbf{Linear Transformations for Queries, Keys, and Values}: For each modality's input $x$ (either $v_i$ or $q_i$), we generate queries $Q$, keys $K$, and values $V$ through linear transformations: 
\begin{equation}
    Q = xW^Q, \quad K = xW^K, \quad V = xW^V 
\end{equation}
\item \textbf{Scaled Dot-Product Attention}: We compute attention scores by taking the dot product of $Q$ and $K$ scaled by $\sqrt{d_k}$, and applying a softmax to emphasize relevant information: 
\begin{equation} 
\text{Attention}(Q, K, V) = \text{softmax}\left(\frac{QK^T}{\sqrt{d_k}}\right)V 
\end{equation}
\item \textbf{Multi-Head Attention Output}: The outputs from multiple attention heads are concatenated and linearly transformed to form the final refined features:
\begin{equation} 
\tilde{x} = \text{Concat}(\text{head}_1, \dots, \text{head}_h)W^O \end{equation}
\end{itemize}

\subsubsection{Cross-modal Bilinear Attention}

After intra-modal refinement, OMniBAN applies a bilinear attention mechanism to fuse the refined visual and textual features. This mechanism captures interactions between modalities by evaluating attention distributions across all pairs of input channels, enabling comprehensive cross-modal feature integration.

To compute the bilinear attention map $ \mathbf{A} $, we use learnable projection matrices $ \mathbf{W}_v \in \mathbb{R}^{d_v \times d_m} $ and $ \mathbf{W}_q \in \mathbb{R}^{d_q \times d_m} $ for the refined image and question features, respectively, where $ d_m $ is the dimension of the shared attention space. The attention map $ \mathbf{A} $ is calculated as:
\begin{equation}
\mathbf{A} = \text{softmax}\left( \left( \mathbf{W}_v \tilde{v}_i \right) \circ \left( \mathbf{W}_q \tilde{q}_i \right) \right)    
\end{equation}

where $ \circ $ denotes the Hadamard (element-wise) product, allowing for efficient alignment between corresponding elements in the image and question features.

The bilinear attention features for each attention head $ h $ are then computed by summing over all interactions between image and question features, weighted by the attention map:
\begin{equation}
\mathbf{f}_h = \sum_{j=1}^{N_v} \sum_{k=1}^{N_q} \mathbf{A}_{jk} \left( \tilde{v}_i^j \right)^T \mathbf{W}_{v, h} \mathbf{W}_{q, h} \tilde{q}_i^k
\end{equation}

In this formulation, each attention glimpse $ h $ learns specialized cross-modal relationships to capture diverse interaction patterns between visual and textual features.

\subsubsection{Orthogonality Loss}

To ensure diverse information captured by the model, we introduce an \textbf{Orthogonality Loss}\cite{yang2020orthogonality} to encourage each attention glimpse to focus on unique aspects of the input. This loss reduces redundancy across glimpses, which is beneficial considering the subtle and often highly similar pixels present in radiological or pathological medical images. Additionally, it helps address with over-fitting during training, which is common in the field of MedVQA.

In detail, we directly apply Orthogonality Loss ($\mathcal{L}_\perp$) to the attention distributions obtained from the bilinear attention, as shown in Algorithm \ref{alg1}. For each pair of attention distributions (glimpses), the inner product of their normalized vectors is computed and squared, with these values summed to form the final orthogonality loss.
\begin{algorithm}
    \caption{Attention Maps with Orthogonality Loss} 
    \begin{algorithmic}[1]
        \Require Visual features $\mathbf{V} \in \mathbb{R}^{B \times N \times D_v}$, Textual features $\mathbf{Q} \in \mathbb{R}^{B \times T \times D_q}$, Number of glimpses $G$, Mask $M$
        
        \State \textbf{Step 1: Compute Attention Scores}
        \State Calculate attention scores: $\mathbf{S} \gets f(\mathbf{V}, \mathbf{Q})$
        \If{$M$ is applied}
            \State Apply mask $M$ to $\mathbf{S}$ to ignore invalid regions
        \EndIf

        \State \textbf{Step 2: Normalize Scores into Distributions}
        \State Compute attention distributions: $\mathbf{P} \gets \text{Softmax}(\mathbf{S})$

        \State \textbf{Step 3: Compute Orthogonality Loss}
        \State $\mathcal{L}_\perp \gets 0$
        \For{each pair of glimpses $(g_1, g_2)$ where $g_1 \neq g_2$}
            \State Normalize distributions $\mathbf{P}_{g_1}$ and $\mathbf{P}_{g_2}$
            \State Compute inner product: $\rho \gets \mathbf{P}_{g_1} \cdot \mathbf{P}_{g_2}$
            \State Update orthogonality loss: $\mathcal{L}_\perp \gets \mathcal{L}_\perp + \rho^2$
        \EndFor
    \end{algorithmic} 
    \label{alg1}
\end{algorithm}

\subsubsection{Classifier and Loss Function}
The joint representation output from BAN is then fed into a classifier to predict the most probable answer. A simple feed-forward neural network is used as the classifier, consisting of two fully connected layers with an intermediate activation function.

The main loss function for this task is Binary Cross-Entropy with Logits Loss, which is a common choice for multi-label classification. The total loss during training combines the main classification loss ($\mathcal{L}_\bullet$) with the Orthogonality Loss ($\mathcal{L}_\perp$), which encourages both accurate predictions and diversified attention features.
\begin{equation}
    \mathcal{L} = \mathcal{L}_\bullet + \alpha \cdot \mathcal{L}_\perp
\end{equation}
where $\alpha$ is the linear threshold for Orthogonality Loss.

\subsubsection{Theoretical Complexity Analysis}
\label{sec:complexity}

\noindent Let $\mathbf{V} \in \mathbb{R}^{N_v \times d_v}$ represent the visual features extracted from $N_v$ image regions ($N_v=1$ in the context of CLIP-based visual backbones), and let $\mathbf{Q} \in \mathbb{R}^{N_q \times d_q}$ denote the textual features for $N_q$ tokens. We analyze and compare the computational complexity of two fusion paradigms: a vanilla Co-Attention mechanism\cite{MCAN}\cite{hcoatt} (which is implemented as a stack of Transformer layers) and our proposed OMniBAN framework.

Both Transformer-based Co-Attention and OMniBAN networks begin by applying intra-modal self-attention to refine the visual and textual features independently. This initial processing step has a computational complexity of:
\begin{equation}
    \mathcal{O}_{\text{self-att}} = \mathcal{O}(N_v^2 d_v + N_q^2 d_q)
\end{equation}

\textbf{Transformer Co-Attention.} A standard Transformer-based co-attention layer processes visual and textual features through cross-modal attention and feed-forward networks (FFN). The dominant operations \textit{per layer} include:
\begin{itemize}
\item Cross-modal Attention: Projecting features and performing attention between sequences of length $N_v$ and $N_q$ with dimensions $d_v$ and $d_q$ results in a cost scaling with:
\begin{equation}
  \mathcal{O}_{\text{cross-att}} = \mathcal{O}(N_q N_v d_q + N_v d_v d_q + N_q d_q^2)  
\end{equation}
where $\mathcal{O}(N_q N_v d_q)$ denotes the core interaction matrix multiplication cost, assuming that attention dimension is comparable to $d_q$.
\item FFN: For a sequence of length $N$ and dimension $d$, this costs $\mathcal{O}(N d^2)$. In the co-attention model, this is typically applied to the output sequences of both modalities, and its cost is:
\begin{equation}
\mathcal{O}_{\text{ffn}} = \mathcal{O}(N_q d_q^2 + N_v d_v^2)    
\end{equation}
\end{itemize}

For a stack of $L$ co-attention layers, the total fusion complexity can be expressed as:
\begin{equation}
\begin{split}
\mathcal{O}_{\text{co-att}} &= \mathcal{O}(N_v^2 d_v + N_q^2 d_q)\\
&+ L \cdot (\mathcal{O}(N_q N_v d_q + N_v d_v d_q + N_q d_q^2 + N_v d_v^2))
\end{split}
\end{equation}
where the dominant terms are:
\begin{equation}
\mathcal{O}(N_v^2 d_v + N_q^2 d_q + L N_q N_v \max(d_v, d_q) + L (N_q d_q^2 + N_v d_v^2))    
\end{equation}

\textbf{OMniBAN Fusion.} Following the same intra-modal self-attention, OMniBAN employs Factorized Bilinear Attention Networks \cite{BAN} across $\gamma$ glimpses. Each glimpse includes:
\begin{itemize} 
    \item Factorized Bilinear Interaction: Computing cross-modal interactions with a cost much lower than $\mathcal{O}(N_v N_q d_v d_q)$ through factorization and projection to a lower intermediate dimension $d_m$ ($d_m \ll \min(d_v, d_q)$), which scales approximately with $\mathcal{O}(N_q N_v d_m)$.
    \item Post-interaction Projection: A simple projection layer after the bilinear step costs $\mathcal{O}(N_q d_q^2)$.
\end{itemize}

Therefore, the total fusion complexity is:
\begin{equation}
\mathcal{O}_{\text{OMniBAN}} = \mathcal{O}(N_v^2 d_v + N_q^2 d_q) + \gamma \cdot (\mathcal{O}(N_q N_v d_m) + \mathcal{O}(N_q d_q^2)))
\end{equation}

\textbf{Complexity Comparison.} Assuming $d_v \approx d_q \approx d$, we can further compare the dominant computational costs of Transformer-based co-attention and OMniBAN. The initial intra-modal self-attention, which costs $\mathcal{O}((N_v^2 + N_q^2) d)$, is common to both frameworks and less dominant than the costs of the repeated fusion steps for $L, \gamma > 1$. When $L \approx \gamma$, the total complexity difference is primarily determined by the cost per repeated step. OMniBAN's step is more efficient due to:
\begin{itemize}
    \item Cross-modal Interaction: Replacing Transformer's attention $\mathcal{O}(N_q N_v d)$ with factorized bilinear $\mathcal{O}(N_q N_v d_m)$, where $d_m \ll d$.
    \item Post-interaction Operation: Using a simpler projection $\mathcal{O}(N_q d^2)$ compared to Transformer's FFN $\mathcal{O}((N_q+N_v) d^2)$. While both exhibit the same asymptotic complexity concerning $N_q$ and $d_q$, the Transformer's FFN involves a significantly larger constant factor (due to an expanded intermediate layer), which may lead to higher practical FLOPs.
\end{itemize}

We also empirically confirm these theoretical efficiency gains in Section~\ref{Experiment}, where we observe that OMniBAN has fewer parameters and reports much smaller FLOPs compared with co-attention.

\section{Experiments}\label{Experiment}

\subsection{Datasets and Metrics}
We conduct our experiments on two public medical VQA datasets: VQA-RAD\cite{VQARAD} and SLAKE\cite{SLAKE}. VQA-RAD contains 3,515 QA pairs derived from 315 radiology images, and is split into 3,064 QA pairs for training and 451 for testing. SLAKE is a Chinese–English bilingual dataset featuring 642 radiology images and a total of 7,033 QA pairs. It includes richer image modalities and covers a broader range of body parts in its questions. For this study, we focus on the English subset of SLAKE (marked as SLAKE-EN in Table \ref{tab-slake}), which consists of 4,919 QA pairs from 450 images in the training set and 1,061 QA pairs from 96 images in the test set. Both VQA-RAD and SLAKE organize their questions into two types: open-ended and closed-ended. Closed-ended questions typically have a limited set of possible responses (most commonly yes/no), while open-ended questions involve more varied answers.

Since MedVQA can be considered as a multi-label classification task, we adopt a binary cross-entropy (BCE) loss function during training. We primarily use accuracy as our evaluation metric, which is computed as the ratio of correctly predicted answers to the total number of questions. To provide a more comprehensive assessment, we report three separate accuracy scores: overall, open-ended, and closed-ended. This breakdown helps compare model performance across different question types.

\subsection{Experimental Setup}
We conduct our experiments on a single NVIDIA Tesla V100-SXM2 (16GB) GPU. The learning rate is set to 0.0005, and the batch size is 32. The number of heads in Multi-head Attention and glimpses in BAN are set to 8 and 5, respectively. For the orthogonality loss, we adopt a strategy that linearly increases its weight throughout training (up to 0.5). We train the models for 40 epochs on both the VQA-RAD dataset and the SLAKE-EN dataset, and save the best-performing model on the validation set as the representative model. Model parameters are optimized using the Adamax optimizer. To mitigate randomness, we train the OMniBAN model ten times with different random seeds and report the average performance along with the standard deviation in Table \ref{tab2}.

At the model level, we integrate the CR approach \cite{QCR}, which uses a pre-trained question classifier to determine whether a given question is open-ended or closed-ended. Based on this classification, the question is routed to one of two specialized sub-models. This design directly addresses the distinct linguistic structures and answer formats in open versus closed questions, which can help the overall model capture the subtle semantic differences of question types more effectively. As shown in Table \ref{tab1} and \ref{tab-slake}, models that leverage this approach are marked with “CR”.

\subsection{Results and Analysis}
\begin{table}
\centering
\caption{Comparison of Accuracy (\%) on VQA-RAD\cite{VQARAD} test set. The highest accuracy in each column is marked in \textbf{bold}. Results are \underline{underlined} to indicate the improvement by OMniBAN.}
\vspace{-4pt}
\resizebox{\linewidth}{!}{
\begin{tabular}{ccccc}
\toprule
\multirow{2}{*}{\textbf{Reference Methods}} & \multirow{2}{*}{\textbf{Fusion Methods}} & \multicolumn{3}{c}{\textbf{Accuracy}} \\
\cmidrule{3-5}
 &  & \textbf{\textit{Open}} & \textbf{\textit{Closed}} & \textbf{\textit{All}} \\
\midrule
MEVF\cite{MEVF} & SAN & 40.7 & 74.1 & 60.8 \\
MEVF\cite{MEVF} & BAN & 43.9 & 75.1 & 62.7 \\
MMQ\cite{MMQ} & BAN & 53.7 & 75.8 & 67.0 \\
CR\cite{QCR} & BAN & 60.0 & 79.3 & 71.6 \\
CPRD\cite{CPRD} & BAN & 52.5 & 77.9 & 67.8\\
\midrule
PubMedCLIP(MEVF)\cite{pubmedclip} & BAN & 48.6 & 78.1 & 66.5 \\
PubMedCLIP(CR)\cite{pubmedclip} & BAN & 60.1 & 80.0 & 72.1 \\
PubMedCLIP(CR)\cite{pubmedclip} & \textbf{OMniBAN(Ours)} & 57.4 & \underline{80.6} & 71.4 \\
\midrule
BiomedCLIP\cite{biomedclip} & Transformer & \textbf{67.6} & 79.8 & \textbf{75.2} \\
BiomedCLIP(CR)\cite{biomedclip} & \textbf{OMniBAN(Ours)} & 66.4 & \underline{\textbf{80.9}} & 75.1 \\
\bottomrule
\end{tabular}
}
\label{tab1}
\end{table}

\begin{table}
\centering
\caption{Comparison of Accuracy (\%) on SLAKE-EN\cite{SLAKE} test set. The highest accuracy in each column is marked in \textbf{bold}. Results are \underline{underlined} to indicate the improvement by OMniBAN.}
\vspace{-4pt}
\begin{threeparttable}
\resizebox{\linewidth}{!}{
\begin{tabular}{ccccc}
\toprule
\multirow{2}{*}{\textbf{Reference Methods}} & \multirow{2}{*}{\textbf{Fusion Methods}} & \multicolumn{3}{c}{\textbf{Accuracy}} \\
\cmidrule{3-5}
 &  & \textbf{\textit{Open}} & \textbf{\textit{Closed}} & \textbf{\textit{All}} \\
\midrule     
MEVF\cite{MEVF} & SAN & 75.3 & 78.4 & 76.5 \\
MEVF\cite{MEVF} & BAN & 77.8 & 79.8 & 78.6 \\
MMQ\IEEEauthorrefmark{2}\cite{MMQ} & BAN & - & - & - \\
CR\cite{QCR} & BAN & 78.8 & 82.0 & 80.0 \\
CPRD\cite{CPRD} & BAN & 79.5 & 83.4 & 81.1\\
\midrule  
PubMedCLIP(MEVF)\cite{pubmedclip} & BAN & 76.5 & 80.4 & 78.0 \\
PubMedCLIP(CR)\cite{pubmedclip} & BAN & 78.4 & 82.5 & 80.1 \\
PubMedCLIP(CR)\cite{pubmedclip} & \textbf{OMniBAN(Ours)} & 78.1 & \underline{85.8} & \underline{81.1} \\
\midrule  
BiomedCLIP\cite{biomedclip} & Transformer & \textbf{82.5} & 89.7 & \textbf{85.4} \\
BiomedCLIP(CR)\cite{biomedclip} & \textbf{OMniBAN(Ours)} & 82.0 & \underline{\textbf{89.9}} & 85.1 \\
\bottomrule
\end{tabular}
}
\begin{tablenotes}
\footnotesize
\item{\IEEEauthorrefmark{2}}MMQ not reported on the SLAKE dataset.
\end{tablenotes}
\end{threeparttable}
\label{tab-slake}
\end{table}

\begin{table*}
\caption{Ablation Study on VQA-RAD and SLAKE-EN test sets (\%).}
\begin{threeparttable}
\resizebox{\linewidth}{!}{
\begin{tabular}{cccccccc}
\toprule
\multirow{2}{*}{\textbf{Dataset}} & \multirow{2}{*}{\textbf{Image Encoder}} & \multirow{2}{*}{\textbf{Text Encoder}} & \multirow{2}{*}{\textbf{Multi-head Attention}} & \multirow{2}{*}{\textbf{Orthogonality Loss}} & \multicolumn{3}{c}{\textbf{Accuracy}} \\
\cmidrule{6-8} 
& & & & & \textbf{\textit{Open}} & \textbf{\textit{Closed}} & \textbf{\textit{All}} \\
\midrule
\multirow{3}{*}{VQA-RAD} & \multirow{3}{*}{BiomedCLIP\cite{biomedclip}} & \multirow{3}{*}{BioBERT\cite{BioBERT}} & - & - & 54.3 ± 3.3 & 77.3 ± 1.9 & 68.2 ± 2.0 \\
&  &  & \checkmark & - & 64.3 ± 1.1 & 79.4 ± 0.9 & 73.4 ± 0.5 \\
&  &  & \checkmark & \checkmark & \textbf{66.4 ± 1.0} & \textbf{80.9 ± 1.3} & \textbf{75.1 ± 0.8} \\
\midrule
\multirow{3}{*}{SLAKE-EN} & \multirow{3}{*}{BiomedCLIP\cite{biomedclip}} & \multirow{3}{*}{BioBERT\cite{BioBERT}} & - & - & 79.0 ± 0.6 & 84.2 ± 1.4 & 81.1 ± 1.1 \\
&  &  & \checkmark & - & 80.6 ± 0.4 & 87.2 ± 1.2 & 83.2 ± 0.5 \\
&  &  & \checkmark & \checkmark & \textbf{82.0 ± 0.2} & \textbf{89.9 ± 1.1} & \textbf{85.1 ± 0.6} \\
\bottomrule
\end{tabular}
}
\label{tab2}
\end{threeparttable}
\end{table*}

\textbf{Comparisons on VQA-RAD Dataset.} Table \ref{tab1} shows our results on the VQA-RAD test set, comparing open-ended, closed-ended, and overall accuracy. The baseline models MEVF+SAN \cite{MEVF} and MEVF+BAN \cite{MEVF} achieve 60.8\% and 62.7\% overall accuracy, respectively, with BAN outperforming SAN due to its bilinear attention mechanism. MMQ \cite{MMQ} further raises performance to 67.0\%, while CR \cite{QCR} brings a notable jump to 71.6\% by classifying questions into open or closed types before prediction. Leveraging PubMedCLIP with BAN leads to 66.5\% accuracy, which increases to 72.1\% when combined with CR. Incorporating OMniBAN improves closed-ended performance to 80.6\% (PubMedCLIP) and 80.9\% (BiomedCLIP), which indicates its strength on questions with restricted answer sets. Although open-ended accuracy dips slightly, overall performance remains competitive, and it demonstrates that OMniBAN can efficiently capture the patterns of closed questions while keeping pace with Transformer-based fusion approaches.

\textbf{Comparisons on SLAKE-EN Dataset.} We observe similar trends on the SLAKE-EN test set in Table \ref{tab-slake}. Again, CR provides a consistent boost over simpler BAN baselines, and OMniBAN excels at closed-ended queries. For instance, PubMedCLIP(CR)+OMniBAN reaches 85.8\% on closed-ended questions and 81.1\% overall, which improves upon the 82.5\% closed-ended accuracy and 80.1\% overall accuracy of PubMedCLIP(CR)+BAN. Likewise, BiomedCLIP(CR)+OMniBAN achieves 89.9\% closed-ended accuracy, slightly above BiomedCLIP’s 89.7\%, but with a marginal decrease in open-ended accuracy.

\begin{table}
\centering
\caption{Comparison of Computational Efficiency on VQA-RAD \cite{VQARAD} Training Set using parameter sizes (M) and FLOPs (M)}
\begin{threeparttable}
\resizebox{0.8\linewidth}{!}{
\begin{tabular}{ccc}
\hline
\textbf{Methods}& \textbf{Co-Attention} & \textbf{OMniBAN} \\
\hline     
Parameters (M) & 31.910 & 21.659 \\
FLOPs (M) & 701.276 & 182.059 \\
\hline
\end{tabular}
}
\end{threeparttable}
\label{tab-efficiency}
\end{table}

\textbf{Efficiency Comparison.} In order to demonstrate OMniBAN’s high efficiency, we also compare a typical Transformer-based fusion method, i.e., Co-Attention \cite{MCAN} \cite{hcoatt} with OMniBAN, as shown in Table \ref{tab-efficiency}. The experiment is conducted on the VQA-RAD training set and evaluated via parameter size and FLOPs. To ensure fairness, the Co-Attention fusion approach includes five layers of cross-modal Transformer encoding (alongside one image–text intra-modal attention layer) to match the five glimpses used by OMniBAN’s bilinear attention. 

The results turn out that OMniBAN requires fewer parameters (21.659M vs.\ 31.910M) and significantly fewer FLOPs (182.059M vs.\ 701.276M), which indicates that its bilinear attention component can effectively reduce computational overhead compared with Transformer-level fusion method. This makes OMniBAN a compelling choice for MedVQA scenarios where efficiency is of priority.

\subsection{Ablation Study}

We further explore the contribution from each component in the OMniBAN. Table \ref{tab2} presents an ablation study on both the VQA-RAD and SLAKE-EN test sets, starting with a baseline model that uses BAN for cross-modal fusion, BiomedCLIP \cite{biomedclip} as the image encoder, and BioBERT \cite{BioBERT} as the text encoder, without multi-head attention or orthogonality loss. On VQA-RAD, this baseline achieves 54.3\% open, 77.3\% closed, and 68.2\% overall accuracy. Introducing multi-head attention boosts overall accuracy to 73.4\%, which reveals the importance of refining intra-modal representations before combining them via BAN. Finally, adding orthogonality loss brings an additional, though smaller, increase to 75.1\%. We attribute this limited incremental gain to the inherent constraints of the encoders, which cap the potential benefits of non-overlapping attention glimpses.

On SLAKE-EN, the same pattern emerges. The baseline obtains 79.0\% open, 84.2\% closed, and 81.1\% overall accuracy. Incorporating multi-head attention again yields a notable jump to 83.2\% overall, while orthogonality loss provides a further marginal improvement. These findings confirm the value of single-modality attention and orthogonality in enhancing BAN, and suggests that given sufficiently robust features from domain-specific encoders, bilinear fusion can perform competitively with Transformer-based methods for MedVQA.

\section{Conclusion}
In this paper, we introduced the Orthogonal Multi-head Bilinear Attention Network (OMniBAN) as an efficient fusion framework for Medical Visual Question Answering. By combining a single-layer multi-head self-attention mechanism with bilinear attention and employing Orthogonality Loss, OMniBAN balances accuracy and computational efficiency. Experimental results on VQA-RAD and SLAKE-EN show that OMniBAN, when paired with BiomedCLIP, slightly surpasses the original Transformer-based BiomedCLIP model on closed-type questions and achieves quite similar overall accuracy. This indicates that bilinear attention is capable of capturing the structured patterns often found in such questions, and offers a promising alternative to Transformer-based fusion methods without sacrificing performance.

Beyond efficiency, OMniBAN’s underlying design has broader implications for visual–textual interaction and model adaptability. Future research could delve more deeply into these aspects by examining OMniBAN’s robustness under different data distributions or its ability to integrate external medical knowledge. Refinements to BAN’s internal attention mechanisms also present an avenue for further enhancing cross-modal interactions. Nevertheless, our work has a few limitations. We have not evaluated OMniBAN on broader and more specialized MedVQA datasets (e.g., Surgical VQA), and we have only tested two encoders—PubMedCLIP and BiomedCLIP—which leaves the model’s performance with other potential encoders unexamined. These constraints underscore the need for broader experimentation and more diverse ablations. Overall, our findings show that OMniBAN can serve as an effective and efficient choice for Medical Visual Question Answering, which indicates that bilinear attention deserves continued exploration in medical image understanding.

\bibliographystyle{IEEEtran}   
\bibliography{IEEEabrv,ref}

\begin{thebibliography}{10}
\providecommand{\url}[1]{#1}
\csname url@samestyle\endcsname
\providecommand{\newblock}{\relax}
\providecommand{\bibinfo}[2]{#2}
\providecommand{\BIBentrySTDinterwordspacing}{\spaceskip=0pt\relax}
\providecommand{\BIBentryALTinterwordstretchfactor}{4}
\providecommand{\BIBentryALTinterwordspacing}{\spaceskip=\fontdimen2\font plus
\BIBentryALTinterwordstretchfactor\fontdimen3\font minus \fontdimen4\font\relax}
\providecommand{\BIBforeignlanguage}[2]{{%
\expandafter\ifx\csname l@#1\endcsname\relax
\typeout{** WARNING: IEEEtran.bst: No hyphenation pattern has been}%
\typeout{** loaded for the language `#1'. Using the pattern for}%
\typeout{** the default language instead.}%
\else
\language=\csname l@#1\endcsname
\fi
#2}}
\providecommand{\BIBdecl}{\relax}
\BIBdecl

\bibitem{vggnet}
K.~Simonyan, ``Very deep convolutional networks for large-scale image recognition,'' \emph{arXiv preprint arXiv:1409.1556}, 2014.

\bibitem{resnet}
K.~He, X.~Zhang, S.~Ren, and J.~Sun, ``Deep residual learning for image recognition,'' in \emph{Proceedings of the IEEE conference on computer vision and pattern recognition}, 2016, pp. 770--778.

\bibitem{chung2014empirical}
J.~Chung, C.~Gulcehre, K.~Cho, and Y.~Bengio, ``Empirical evaluation of gated recurrent neural networks on sequence modeling,'' \emph{arXiv preprint arXiv:1412.3555}, 2014.

\bibitem{hochreiter1997long}
S.~Hochreiter, ``Long short-term memory,'' \emph{Neural Computation MIT-Press}, 1997.

\bibitem{MEVF}
B.~D. Nguyen, T.-T. Do, B.~X. Nguyen, T.~Do, E.~Tjiputra, and Q.~D. Tran, ``Overcoming data limitation in medical visual question answering,'' in \emph{Medical Image Computing and Computer Assisted Intervention--MICCAI 2019: 22nd International Conference, Shenzhen, China, October 13--17, 2019, Proceedings, Part IV 22}.\hskip 1em plus 0.5em minus 0.4em\relax Springer, 2019, pp. 522--530.

\bibitem{CPRD}
B.~Liu, L.-M. Zhan, and X.-M. Wu, ``Contrastive pre-training and representation distillation for medical visual question answering based on radiology images,'' in \emph{Medical Image Computing and Computer Assisted Intervention--MICCAI 2021: 24th International Conference, Strasbourg, France, September 27--October 1, 2021, Proceedings, Part II 24}.\hskip 1em plus 0.5em minus 0.4em\relax Springer, 2021, pp. 210--220.

\bibitem{SAN}
Z.~Yang, X.~He, J.~Gao, L.~Deng, and A.~Smola, ``Stacked attention networks for image question answering,'' in \emph{Proceedings of the IEEE conference on computer vision and pattern recognition}, 2016, pp. 21--29.

\bibitem{MFB}
Z.~Yu, J.~Yu, J.~Fan, and D.~Tao, ``Multi-modal factorized bilinear pooling with co-attention learning for visual question answering,'' in \emph{Proceedings of the IEEE international conference on computer vision}, 2017, pp. 1821--1830.

\bibitem{MUTAN}
H.~Ben-Younes, R.~Cadene, M.~Cord, and N.~Thome, ``Mutan: Multimodal tucker fusion for visual question answering,'' in \emph{Proceedings of the IEEE international conference on computer vision}, 2017, pp. 2612--2620.

\bibitem{BAN}
J.-H. Kim, J.~Jun, and B.-T. Zhang, ``Bilinear attention networks,'' \emph{Advances in neural information processing systems}, vol.~31, 2018.

\bibitem{transformer}
A.~Vaswani, ``Attention is all you need,'' \emph{arXiv preprint arXiv:1706.03762}, 2017.

\bibitem{lxmert}
H.~Tan and M.~Bansal, ``Lxmert: Learning cross-modality encoder representations from transformers,'' \emph{arXiv preprint arXiv:1908.07490}, 2019.

\bibitem{radstruct}
C.~Pellegrini, M.~Keicher, E.~{\"O}zsoy, and N.~Navab, ``Rad-restruct: A novel vqa benchmark and method for structured radiology reporting,'' in \emph{International Conference on Medical Image Computing and Computer-Assisted Intervention}.\hskip 1em plus 0.5em minus 0.4em\relax Springer, 2023, pp. 409--419.

\bibitem{MCAN}
Z.~Yu, J.~Yu, Y.~Cui, D.~Tao, and Q.~Tian, ``Deep modular co-attention networks for visual question answering,'' in \emph{Proceedings of the IEEE/CVF conference on computer vision and pattern recognition}, 2019, pp. 6281--6290.

\bibitem{METER}
Z.-Y. Dou, Y.~Xu, Z.~Gan, J.~Wang, S.~Wang, L.~Wang, C.~Zhu, P.~Zhang, L.~Yuan, N.~Peng \emph{et~al.}, ``An empirical study of training end-to-end vision-and-language transformers,'' in \emph{Proceedings of the IEEE/CVF Conference on Computer Vision and Pattern Recognition}, 2022, pp. 18\,166--18\,176.

\bibitem{q2a}
Y.~Liu, Z.~Wang, D.~Xu, and L.~Zhou, ``Q2atransformer: Improving medical vqa via an answer querying decoder,'' in \emph{International Conference on Information Processing in Medical Imaging}.\hskip 1em plus 0.5em minus 0.4em\relax Springer, 2023, pp. 445--456.

\bibitem{biomedclip}
S.~Zhang, Y.~Xu, N.~Usuyama, H.~Xu, J.~Bagga, R.~Tinn, S.~Preston, R.~Rao, M.~Wei, N.~Valluri \emph{et~al.}, ``Biomedclip: a multimodal biomedical foundation model pretrained from fifteen million scientific image-text pairs,'' \emph{arXiv preprint arXiv:2303.00915}, 2023.

\bibitem{CLIP}
A.~Radford, J.~W. Kim, C.~Hallacy, A.~Ramesh, G.~Goh, S.~Agarwal, G.~Sastry, A.~Askell, P.~Mishkin, J.~Clark \emph{et~al.}, ``Learning transferable visual models from natural language supervision,'' in \emph{International conference on machine learning}.\hskip 1em plus 0.5em minus 0.4em\relax PMLR, 2021, pp. 8748--8763.

\bibitem{BioBERT}
J.~Lee, W.~Yoon, S.~Kim, D.~Kim, S.~Kim, C.~H. So, and J.~Kang, ``Biobert: a pre-trained biomedical language representation model for biomedical text mining,'' \emph{Bioinformatics}, vol.~36, no.~4, pp. 1234--1240, 2020.

\bibitem{yang2020orthogonality}
S.~Yang, W.~Deng, M.~Wang, J.~Du, and J.~Hu, ``Orthogonality loss: Learning discriminative representations for face recognition,'' \emph{IEEE Transactions on Circuits and Systems for Video Technology}, vol.~31, no.~6, pp. 2301--2314, 2020.

\bibitem{hcoatt}
J.~Lu, J.~Yang, D.~Batra, and D.~Parikh, ``Hierarchical question-image co-attention for visual question answering,'' \emph{Advances in neural information processing systems}, vol.~29, 2016.

\bibitem{VQARAD}
J.~J. Lau, S.~Gayen, A.~Ben~Abacha, and D.~Demner-Fushman, ``A dataset of clinically generated visual questions and answers about radiology images,'' \emph{Scientific data}, vol.~5, no.~1, pp. 1--10, 2018.

\bibitem{SLAKE}
B.~Liu, L.-M. Zhan, L.~Xu, L.~Ma, Y.~Yang, and X.-M. Wu, ``Slake: A semantically-labeled knowledge-enhanced dataset for medical visual question answering,'' in \emph{2021 IEEE 18th International Symposium on Biomedical Imaging (ISBI)}.\hskip 1em plus 0.5em minus 0.4em\relax IEEE, 2021, pp. 1650--1654.

\bibitem{QCR}
L.-M. Zhan, B.~Liu, L.~Fan, J.~Chen, and X.-M. Wu, ``Medical visual question answering via conditional reasoning,'' in \emph{Proceedings of the 28th ACM International Conference on Multimedia}, 2020, pp. 2345--2354.

\bibitem{MMQ}
T.~Do, B.~X. Nguyen, E.~Tjiputra, M.~Tran, Q.~D. Tran, and A.~Nguyen, ``Multiple meta-model quantifying for medical visual question answering,'' in \emph{Medical Image Computing and Computer Assisted Intervention--MICCAI 2021: 24th International Conference, Strasbourg, France, September 27--October 1, 2021, Proceedings, Part V 24}.\hskip 1em plus 0.5em minus 0.4em\relax Springer, 2021, pp. 64--74.

\bibitem{pubmedclip}
S.~Eslami, C.~Meinel, and G.~De~Melo, ``Pubmedclip: How much does clip benefit visual question answering in the medical domain?'' in \emph{Findings of the Association for Computational Linguistics: EACL 2023}, 2023, pp. 1181--1193.

\end{thebibliography}

\end{document}